\documentclass[reprint,amsmath,amssymb,aps,prd,showkeys]{revtex4-2}
\usepackage{graphicx}
\usepackage{dcolumn}
\usepackage{bm}
\usepackage{xcolor}
\usepackage{orcidlink}
\usepackage{hyperref}
\usepackage{tikz}
\usetikzlibrary{shapes.geometric, arrows.meta, positioning, calc, decorations.pathreplacing}

\definecolor{boxborder}{RGB}{44, 110, 157}
\definecolor{boxfill}{RGB}{229, 240, 248}
\definecolor{fgborder}{RGB}{176, 96, 40}
\definecolor{fgfill}{RGB}{250, 236, 222}

\hypersetup{hidelinks}

\begin{document}
\raggedbottom
\setlength{\parskip}{0pt}

\title{Scaling and Cross-Simulation Generalization of Graph Neural Networks for Cosmic-Web Classification in CAMELS}

\author{Thakkar Kalash D.\,\orcidlink{0009-0006-6888-457X}}
\email{msc2503121005@iiti.ac.in}
\affiliation{Department of Astronomy, Astrophysics \& Space Engineering, Indian Institute of Technology Indore, MP 453552, India}

\author{Unnati Shah}
\email{unshah@utica.edu}
\affiliation{Department of Computer Science, Utica University, Utica, NY 13502, USA}

\begin{abstract}
Graph-based methods can classify galaxies by their dark-matter cosmic-web environments, but reliability beyond development simulations remains uncertain. Galaxy-formation models alter tracer populations and spatial distributions, potentially shifting graph features. We test how graph neural networks scale with independent training realizations and transfer across hydrodynamical simulations. Using the $z=0$ CAMELS IllustrisTNG, SIMBA and Astrid suites, we label galaxies with $M_\star>10^9\,M_\odot$ as void, sheet, filament or knot from the smoothed cold-dark-matter density field. We represent their spatial distributions as periodic Delaunay graphs with ten geometric descriptors. We compare a multilayer perceptron (MLP) and five graph neural networks (GNNs), including a local graph Transformer, under realization-grouped evaluation. Learning curves increase training data from 10 to 100 realizations, with validation and test boxes fixed. We test every source-to-target suite pairing on 100 unseen realizations per target suite. On this external holdout, the Transformer, GAT+ and EdgeConv reach mean per-realization macro-averaged F1 scores of $0.652$, $0.651$ and $0.642$, against $0.481$ for the MLP. The Transformer outperforms the MLP by 0.171, while the difference between the Transformer and GAT+ is not statistically significant. Voids remain the hardest class. Performance is higher in boxes with more selected galaxies and shorter Delaunay edges. Associations with cosmological and feedback parameters weaken after accounting for galaxy richness and void fraction. Our results benchmark the scaling and cross-simulation transfer of galaxy-graph classifiers, demonstrating the value of neighbourhood information and identifying tracer sampling as an important limitation for dark-matter web inference.

\end{abstract}


\maketitle

\section{Introduction}
\label{sec:introduction}

On scales of a few to tens of megaparsecs, matter is arranged in a connected network of underdense voids, planar sheets (or walls), filaments and dense knots. Several schemes classify this network from the matter distribution. The tidal web (T-web) counts the collapsing directions of the Hessian of the gravitational potential \citep{Hahn2007,ForeroRomero2009}. The V-web applies the same counting to the velocity shear tensor and resolves structure on scales of $\sim0.1\,h^{-1}\mathrm{Mpc}$, an order of magnitude finer than the tidal web \citep{Hoffman2012}. Topological methods such as \textsc{DisPerSE} identify voids, walls, filaments and nodes from the persistence of critical features in a density field estimated on the Delaunay tessellation of discrete tracers \citep{Sousbie2011}.

These environmental classifications capture information beyond the local matter density. In N-body simulations, the formation times, shapes, and spins of dark-matter haloes depend on their location within the cosmic web, with halo properties differing systematically between knots, filaments, sheets, and voids \citep{Hahn2007}. The influence of the large-scale environment can also extend to observable galaxy properties. For example, in IllustrisTNG100, the probability of misalignment between stellar and gas rotation at z=0 has been linked to the earlier spin of the host dark-matter halo \citep{Duckworth2020}. Assigning galaxies to their underlying tidal-web environments therefore provides a way to connect observable galaxy properties with the anisotropic gravitational structures in which their host haloes formed.

However, dynamical classifications such as the T-web require access to the underlying matter field, whereas observational galaxy surveys provide only a sparse and biased set of tracers. The abundance and spatial clustering of these tracers depend on galaxy formation and baryonic processes. Inferring a dark-matter-defined environment from galaxy positions is consequently an inverse problem: the reliability of the inference depends both on how densely galaxies sample the cosmic web and on how their spatial distribution relates to the underlying matter field. This makes the representation of galaxy positions an important part of the classification problem.

Graph-based representations offer a natural way to describe the spatial organization of discrete galaxy samples. In such representations, galaxies are treated as vertices and connections between neighbouring galaxies encode their local spatial relationships. The resulting graph captures geometric and neighbourhood information that may be difficult to represent using individual galaxy features alone. Previous studies have demonstrated the potential of this approach for cosmic-web classification. \citet{SuarezPerez2021} constructed $\beta$-skeleton graphs from IllustrisTNG galaxies and used tree-based classifiers to predict tidal-web environments, obtaining an accuracy of 74 per cent and a weighted F1 score of $0.728$. \citet{Kololgi2026} used Delaunay triangulations of stellar-mass-selected IllustrisTNG-300 galaxies and trained a graph attention network with edge features (GAT+), achieving an accuracy of 85 per cent for galaxies with $M_\star>10^9\,M_\odot$. Their graph-based model outperformed both a multilayer perceptron (MLP) and a graph convolutional network.

Although these studies establish that galaxy graphs can provide useful information for cosmic-web classification, they leave an important question unresolved: does a classifier trained on one galaxy-formation model retain its predictive ability when applied to a different model? Previous demonstrations have primarily focused on individual simulation suites, with training and test samples divided at the galaxy level. Such evaluations do not establish whether a learned relationship generalizes to independent cosmological realizations, where galaxies occupy different large-scale environments and share different underlying matter fields.

This question is particularly important because different galaxy-formation models can produce different galaxy populations and spatial distributions. Changes in stellar masses, galaxy abundances, and the populations tracing sheets and filaments can alter the geometric properties of the resulting galaxy graphs. A classifier may therefore encounter a shifted distribution of graph features when applied to a simulation generated with a different hydrodynamical code or subgrid-physics prescription. Previous work using graph neural networks (GNNs) with CAMELS galaxy catalogues has investigated related generalization challenges. GNNs trained to infer halo masses from member-galaxy properties retain some accuracy when transferred between simulations with different codes and subgrid models \citep{VillanuevaDomingo2022}. In contrast, GNNs trained to infer the cosmological parameter $\Omega_{\rm m}$ from galaxy positions experience a loss in accuracy under changes in subgrid physics, although unsupervised domain adaptation can improve transfer between IllustrisTNG and SIMBA \citep{VillanuevaDomingo2022Graphs,Roncoli2023}. \citet{Ni2023} further showed that the Astrid suite spans a broader range of galaxy populations than IllustrisTNG and SIMBA, motivating the use of diverse galaxy-formation models when developing methods intended to generalize to unseen simulations. However, these studies focus on global cosmological or halo-level inference rather than the transferability of a per-galaxy cosmic-web classifier across galaxy-formation models.

The Cosmology and Astrophysics with Machine Learning Simulations (CAMELS) project provides a suitable framework for investigating this problem \citep{VillaescusaNavarro2021,Ni2023}. Its Latin-hypercube (LH) sets contain 1,000 simulations per galaxy-formation model, each covering a volume of $(25\,h^{-1}\mathrm{Mpc})^3$, with different initial conditions and variations in cosmological parameters, including $\Omega_{\rm m}$, $\sigma_8$, as well as four feedback parameters. The IllustrisTNG, SIMBA, and Astrid suites have corresponding box sizes and resolutions, and the corresponding simulation products include both galaxy catalogues and gridded matter fields \citep{VillaescusaNavarro2022CMD}. This combination makes it possible to evaluate galaxy-based predictions against independently defined dark-matter environments while holding out entire simulation realizations. Holding out complete realizations is particularly important because galaxies within the same simulation box share the same large-scale matter field and are not independent test samples.

In this work, we investigate the scaling and cross-simulation generalization of graph neural networks for galaxy-level tidal-web classification using the (z=0) CAMELS IllustrisTNG, SIMBA, and Astrid suites. We address three related questions. First, how does classification performance change as the number of independent source-training realizations increases while the validation and test samples remain fixed? Second, how reliably do classifiers transfer between galaxy-formation models and to independent realizations that were not used during training, validation, or model selection? Third, which properties of the galaxy samples and their associated graphs are linked to classification performance? To address these questions, we compare six neural-network architectures using a common graph representation and a realization-grouped evaluation protocol. We conduct learning-curve experiments in which the number of source-training realizations increases from 10 to 100. We then evaluate frozen models on an independent external sample comprising realizations LH100–LH199 for each simulation suite. Finally, we investigate associations between box-level classification performance, galaxy abundance, environmental class fractions, Delaunay graph geometry, and CAMELS simulation parameters. These analyses allow us to assess both the benefits and limitations of graph-based tidal-web inference under changes in galaxy sampling and galaxy-formation physics. We treat the relationships between simulation parameters and performance as statistical associations rather than evidence of causal effects.

The remainder of this paper is organized as follows. Section~\ref{sec:theory} describes the tidal-web classification and the graph descriptors. Section~\ref{sec:simulations} introduces the CAMELS simulations, galaxy selection, and data splits. Section~\ref{sec:methods} presents the neural-network architectures, preprocessing, and evaluation procedures. Section~\ref{sec:analysis} reports the scaling and cross-simulation generalization results, while Section~\ref{sec:discussion} discusses their implications and limitations. Section~\ref{sec:conclusion} summarizes our conclusions. Exploratory analyses of the CAMELS parameter associations are presented in Appendix~\ref{app:exploratory}.

\section{Tidal-web classification and galaxy graphs}
\label{sec:theory}

\subsection{The tidal web}
\label{subsec:tweb}

The classification target is defined from the matter field alone, independently of the graph classifier, using the tidal-web (T-web) prescription \citep{Hahn2007,ForeroRomero2009}. Given the dimensionless overdensity $\delta(\mathbf{x})=\rho(\mathbf{x})/\bar\rho-1$, we solve the normalized Poisson equation
\begin{equation}
\nabla^2\phi(\mathbf{x})=\delta(\mathbf{x})
\label{eq:poisson}
\end{equation}
in the periodic box, where $\phi$ is the peculiar gravitational potential divided by $4\pi G\bar\rho$ (in comoving coordinates). The tidal tensor is the Hessian of this potential,
\begin{equation}
T_{ij}(\mathbf{x})=\frac{\partial^2\phi}{\partial x_i\,\partial x_j},
\qquad \mathrm{tr}\,T=\delta .
\label{eq:tidal}
\end{equation}
In practice the tensor is evaluated in Fourier space. The overdensity is first smoothed with a Gaussian window of radius $R_{\rm s}$,
\begin{equation}
\tilde T_{ij}(\mathbf{k})=\frac{k_ik_j}{k^2}\,\tilde\delta(\mathbf{k})\,
\exp\!\left(-\frac{k^2R_{\rm s}^2}{2}\right),
\qquad \mathbf{k}\neq\mathbf{0},
\label{eq:tidal_fourier}
\end{equation}
and transformed back to the grid. We use $R_{\rm s}=2\,h^{-1}\mathrm{Mpc}$. On the $128^3$ grids of the $25\,h^{-1}\mathrm{Mpc}$ boxes described in Section~\ref{sec:simulations}, a cell is $0.195\,h^{-1}\mathrm{Mpc}$ wide, so the smoothing kernel spans about ten cells.

The three real eigenvalues $\lambda_1\ge\lambda_2\ge\lambda_3$ of $T_{ij}$ describe the local tidal field along its principal axes. A positive eigenvalue indicates compression of the matter distribution along the corresponding axis. Each grid cell is classified by the number of eigenvalues above a fixed threshold $\lambda_{\rm th}$: zero for a void, one for a sheet, two for a filament and three for a knot. \citet{Hahn2007} motivated the case $\lambda_{\rm th}=0$ from the local stability of test-particle orbits, in close analogy with the Zel'dovich approximation. \citet{ForeroRomero2009} showed that a positive threshold gives a more realistic web: $\lambda_{\rm th}=0.1$ already reproduces the visual impression of the cosmic web \citep{Kololgi2026}, and for $0.2\lesssim\lambda_{\rm th}\lesssim0.4$ a system of percolating voids coexists with a connected network of filaments. We adopt $\lambda_{\rm th}=0.2$, at the lower end of that range. Galaxy-based classifiers are sensitive to this choice: for $\beta$-skeleton features and a $2$\,Mpc smoothing, \citet{SuarezPerez2021} obtained the closest match to the T-web with $\lambda_{\rm th}=0$. The smoothing length and eigenvalue threshold define the classification task here. They were fixed in advance and were not tuned on any test set.

A galaxy takes the label of the classified grid at its position. For a periodic box of side length $L$ resolved on
an $N^3$ grid, the cell index along each coordinate is
\begin{equation}
    i_\alpha =
    \left\lfloor \frac{N x_\alpha}{L} \right\rfloor
    \bmod N,
    \qquad \alpha \in \{x,y,z\}.
\end{equation}
Galaxy positions are wrapped into the periodic domain before
indexing. No interpolation between neighbouring cells is performed. The velocity-shear web of \citet{Hoffman2012} would resolve finer structure, down to $\sim0.1\,h^{-1}\mathrm{Mpc}$, but it requires a well-sampled velocity field. The tidal web needs only the density and is better matched to the scales probed by a sparse galaxy sample.

\begin{figure*}
\centering
\includegraphics[width=0.92\textwidth]{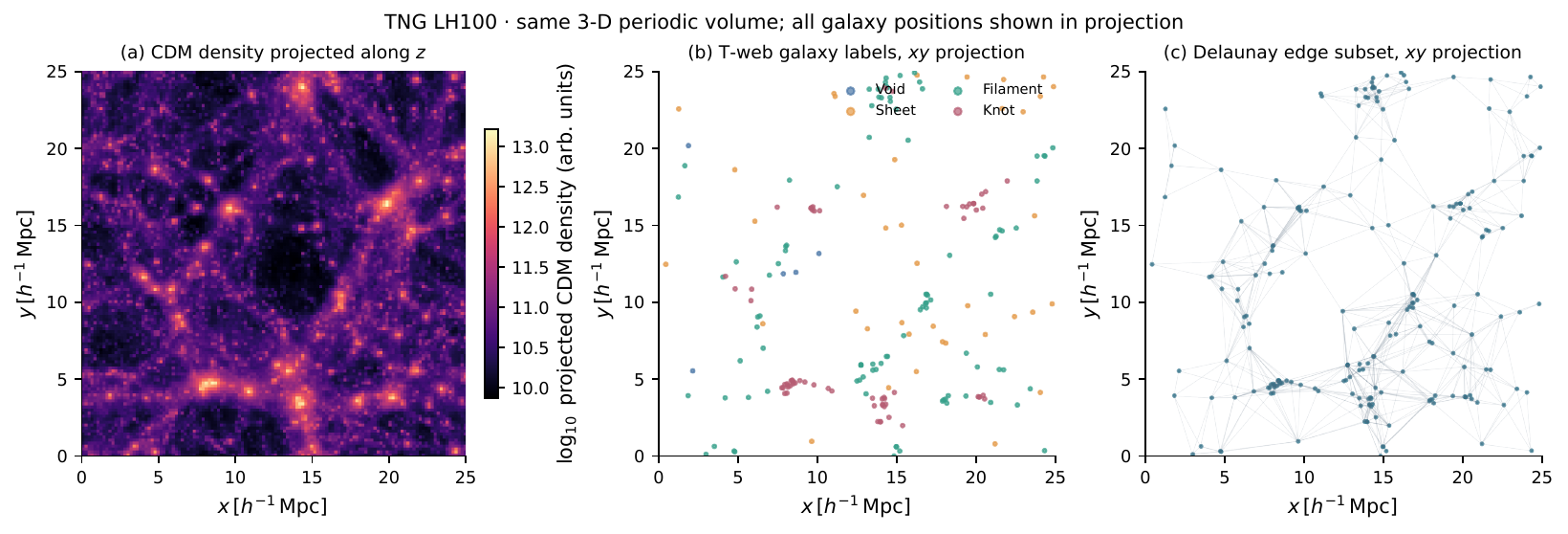}
\caption{An illustrative CAMELS simulation volume (IllustrisTNG, realization LH100). (a) Cold-dark-matter density projected along $z$. (b) T-web labels of the selected galaxies. (c) A subset of the Delaunay edges between the same galaxies. All panels show projections of one periodic three-dimensional volume. The graph is built from galaxy positions alone; the dark-matter field supplies the reference T-web labels used for training and evaluation.}
\label{fig:pipeline}
\end{figure*}

\subsection{Periodic Delaunay graphs}
\label{subsec:graph}

Each realization yields a graph $G=(V,E)$ whose vertices are the galaxies that pass the stellar-mass selection. The edges come from the Delaunay triangulation of the galaxy positions: the decomposition of the volume into tetrahedra whose circumscribed spheres contain no other galaxy. Two galaxies are connected if they share a tetrahedron edge. The triangulation adapts to the local tracer density, so dense regions are linked by short edges and voids by long ones, without a linking length fixed across simulations with different galaxy abundances. The same tessellation underlies density estimators and topological web finders for discrete tracers \citep{Sousbie2011}.

\citet{Kololgi2026} built a non-periodic triangulation of the $205\,h^{-1}\mathrm{Mpc}$ IllustrisTNG-300 box and excluded galaxies within a $10$\,Mpc buffer of each face, where elongated sliver tetrahedra distort the graph metrics. A buffer of that width would remove most of a $25\,h^{-1}\mathrm{Mpc}$ CAMELS volume, so we triangulate with periodic boundary conditions instead. 
Periodic boundaries are incorporated by replicating the galaxy
positions into a $3\times3\times3$ arrangement of neighbouring box images. A three-dimensional Delaunay triangulation is then
constructed using the SciPy implementation of Qhull.
Edges belonging to simplices with at least one vertex in the
central image are mapped back to their original galaxy indices,
and duplicate undirected edges are removed.

Edge lengths are evaluated using the minimum-image convention,
\begin{equation}
    \Delta \mathbf{r}_{ij} =
    \left[
    (\mathbf{r}_j-\mathbf{r}_i+L/2)\bmod L
    \right]-L/2,
    \qquad
    \ell_{ij}=\left|\Delta\mathbf{r}_{ij}\right|.
\end{equation}
This construction preserves connections across periodic
boundaries without excluding galaxies near the box faces. All separations use the minimum-image convention, and each edge carries the periodic separation $\ell_{ij}=|\Delta\mathbf{r}_{ij}|$ of its two galaxies as an attribute.

Every galaxy carries the same ten descriptors in all experiments, following the definitions of \citet{Kololgi2026}; they are listed in Table~\ref{tab:features}. 
Let $\mathcal{N}(i)$ denote the Delaunay neighbours of galaxy
$i$, and let $k_i=|\mathcal{N}(i)|$ be its unweighted degree.
The first input descriptor is the edge-length-weighted degree,
or strength,
\begin{equation}
    s_i=\sum_{j\in\mathcal{N}(i)}\ell_{ij}.
    \label{eq:edgelength}
\end{equation}
The unweighted degree $k_i$ is retained for the normalization
of the clustering coefficient, mean edge length and tetrahedral
density. The model input uses $s_i$ rather than $k_i$ as its
first descriptor.
The clustering coefficient uses edge lengths normalized by the longest edge in the graph, $\hat w_{ij}=\ell_{ij}/\max(\ell)$ for connected pairs and zero otherwise,
\begin{equation}
c_i=\frac{1}{k_i(k_i-1)}\sum_{j,m\in\mathcal{N}(i)}\left(\hat w_{ij}\,\hat w_{im}\,\hat w_{jm}\right)^{1/3}.
\label{eq:clustering}
\end{equation}
Three descriptors summarise the edges attached to each galaxy,
\begin{equation}
\bar\ell_i=\frac{1}{k_i}\sum_{j\in\mathcal{N}(i)}\ell_{ij},\quad
\ell_i^{\min}=\min_{j\in\mathcal{N}(i)}\ell_{ij},\quad
\ell_i^{\max}=\max_{j\in\mathcal{N}(i)}\ell_{ij}.
\label{eq:edges}
\end{equation}

Two further descriptors characterize the local Delaunay
tetrahedra. Let $\mathcal{T}_i$ denote the set of valid
tetrahedra incident on galaxy $i$, let
$T_i=|\mathcal{T}_i|$, and let $V_t$ be the volume of
tetrahedron $t$. We define the tetrahedral density as
\begin{equation}
    \rho_{\mathrm{tet},i}
    =
    \frac{T_i}
    {k_i \displaystyle\sum_{t\in\mathcal{T}_i}V_t},
\end{equation}
where $k_i$ is the unweighted number of Delaunay neighbours.

The neighbouring tetrahedral-density descriptor is the
arithmetic mean of $\rho_{\mathrm{tet}}$ over the Delaunay
neighbours of galaxy $i$, excluding undefined values.
These quantities describe the spatial geometry of the
galaxy distribution and do not use the dark-matter density
field. 
The last three are the ordered eigenvalues $I_1\le I_2\le I_3$ of the covariance matrix of neighbour positions,
\begin{equation}
\mathsf{C}_i=\frac{1}{k_i}\sum_{j\in\mathcal{N}(i)}(\mathbf{r}_j-\bar{\mathbf{r}}_i)(\mathbf{r}_j-\bar{\mathbf{r}}_i)^{\mathsf T},
\qquad
\bar{\mathbf{r}}_i=\frac{1}{k_i}\sum_{j\in\mathcal{N}(i)}\mathbf{r}_j .
\label{eq:shape}
\end{equation}

For this covariance calculation, neighbour positions are unwrapped relative to galaxy $i$ using the minimum-image convention before evaluating $\bar{\mathbf{r}}_i$ and $\mathsf{C}_i$.

\begin{table}
\caption{The ten node descriptors computed on the periodic Delaunay graph, defined in Eqs.~(\ref{eq:edgelength})--(\ref{eq:shape}) following \citet{Kololgi2026}.}
\label{tab:features}
\centering
\small
\setlength{\tabcolsep}{4pt}
\begin{tabular}{lll}
\hline
Descriptor & Symbol & Information \\
\hline
Edge length weighted degree & $s_i$ & connectivity \\
Clustering coefficient & $c_i$ & clustering \\
Mean edge length & $\bar\ell_i$ & geometry \\
Minimum edge length & $\ell_i^{\min}$ & geometry \\
Maximum edge length & $\ell_i^{\max}$ & geometry \\
Tetrahedral density & $\rho_{{\rm tet},i}$ & local density \\
Neighbour tetrahedral density & $\langle\rho_{\rm tet}\rangle_{\mathcal{N}(i)}$ & local density \\
Shape eigenvalues & $I_1,I_2,I_3$ & morphology \\
\hline
\end{tabular}
\end{table}

The shape eigenvalues measure the anisotropy of the neighbourhood. Roughly equal values describe an isotropic configuration, $I_1\ll I_2\approx I_3$ a flattened one and $I_1\approx I_2\ll I_3$ an elongated one \citep{Kololgi2026}. The implementation and feature ordering are held fixed throughout the benchmarks.

The descriptors are computed from the positions of the selected galaxies, whereas the labels come from the eigenvalues of the smoothed dark-matter field. The network must therefore infer the matter-web environment from the geometry of its galaxy tracers, and neither the labels nor the dark-matter field enter its input.

\section{Simulations and samples}
\label{sec:simulations}

\subsection{CAMELS suites and density fields}
\label{subsec:camels}

We use the LH sets of three CAMELS hydrodynamical suites: IllustrisTNG, run with the moving-mesh code \textsc{arepo}; SIMBA, run with the meshless code \textsc{gizmo} \citep{VillaescusaNavarro2021}; and Astrid, run with the smoothed-particle code \textsc{MP-Gadget} \citep{Ni2023}. Every simulation follows $256^3$ dark-matter particles and $256^3$ initial gas elements in a periodic comoving cube of side $25\,h^{-1}\mathrm{Mpc}$. The cosmological parameters $\Omega_{\rm b}=0.049$, $h=0.6711$ and $n_{\rm s}=0.9624$ are fixed in all runs \citep{VillaescusaNavarro2021}.

Each LH set contains 1000 simulations in which $\Omega_{\rm m}$, $\sigma_8$, two supernova-feedback parameters ($A_{\rm SN1}$, $A_{\rm SN2}$), two AGN-feedback parameters ($A_{\rm AGN1}$, $A_{\rm AGN2}$) and the initial random seed all change from one simulation to the next. The cosmological parameters span $\Omega_{\rm m}\in[0.1,0.5]$ and $\sigma_8\in[0.6,1.0]$. The feedback parameters span $[0.25,4]$ for $A_{\rm SN1}$ and $A_{\rm AGN1}$ and $[0.5,2]$ for $A_{\rm SN2}$; $A_{\rm AGN2}$ spans $[0.5,2]$ in IllustrisTNG and SIMBA and $[0.25,4]$ in Astrid \citep{VillaescusaNavarro2021,Ni2023}. The parameter ranges are shared, but the physical processes they modulate differ between galaxy-formation models. In Astrid, for example, $A_{\rm SN2}$ scales the supernova wind speed and $A_{\rm AGN2}$ the efficiency of thermal AGN feedback \citep{Ni2023}, whereas in IllustrisTNG and SIMBA $A_{\rm AGN2}$ sets the ejection speed of kinetic-mode or jet-mode AGN feedback \citep{Ni2023}. The LH sets of different suites also sample different points of the hypercube, so a given index such as LH100 refers to different parameter values in each suite.

We analyse the $z=0$ friends-of-friends and SUBFIND
catalogues \citep{VillaescusaNavarro2021}, stored as \texttt{groups\_090.hdf5},
together with the cold-dark-matter mass-density fields
($M_{\rm cdm}$) from the CAMELS Multifield Dataset
(CMD) \citep{VillaescusaNavarro2022CMD}. For each of the IllustrisTNG, SIMBA and Astrid
Latin-hypercube suites, we use the $z=0$ fields on
periodic $128^3$ grids. Each corresponding LH grid
array contains 1000 realizations, and the grid indexed
by $r$ is paired with the SUBFIND catalogue of
realization LH$r$ from the same simulation suite.
The grid-cell width is
$25/128 \simeq 0.195\,h^{-1}\,\mathrm{Mpc}$.

Galaxies are selected using the stellar-mass component
\texttt{SubhaloMassType[:,4]} of the SUBFIND catalogues.
The stored values $m_{\star,\mathrm{raw}}$ are in units of
$10^{10}\,h^{-1}M_\odot$. To impose the physical
stellar-mass threshold $M_\star>10^9\,M_\odot$, we use
the catalogue's \texttt{HubbleParam} value and require
\begin{equation}
    M_\star =
    \frac{10^{10}}{h}\,
    m_{\star,\mathrm{raw}}\,M_\odot
    > 10^9\,M_\odot.
\end{equation}
For $h=0.6711$, this is equivalent to
$m_{\star,\mathrm{raw}}>0.06711$. In the implementation,
the threshold is converted to the catalogue mass units
before selection. The same stellar-mass criterion is
applied in all three simulation suites.
Positions are wrapped periodically. The number of selected galaxies varies with realization and suite, which motivates the box-level diagnostics of Section~\ref{subsec:physical_analysis}. We make no attempt to match individual galaxies between suites.

\subsection{Development, scaling and external test sets}
\label{subsec:splits}

\begin{figure}
\centering
\resizebox{\columnwidth}{!}{%
\begin{tikzpicture}[font=\footnotesize,
  seg/.style={draw=boxborder, fill=boxfill, minimum height=0.65cm, inner sep=1pt, anchor=west, align=center},
  ext/.style={draw=fgborder, fill=fgfill, minimum height=0.65cm, inner sep=1pt, anchor=west, align=center},
  brc/.style={decorate, decoration={brace, amplitude=3pt}, draw=black!70}]
\node[seg, minimum width=1.7cm] (a) at (0,0) {LH0--9};
\node[seg, minimum width=1.05cm, fill=boxfill!55] (b) at (a.east) {10--29};
\node[seg, minimum width=1.05cm, fill=boxfill!55] (c) at (b.east) {30--49};
\node[seg, minimum width=1.4cm, fill=boxfill!55] (d) at (c.east) {50--99};
\node[ext, minimum width=1.9cm] (e) at ([xshift=2mm]d.east) {LH100--199};
\draw[brc] ([yshift=1mm]a.north west) -- ([yshift=1mm]a.north east) node[midway, above=2pt] {10};
\draw[brc] ([yshift=5.5mm]a.north west) -- ([yshift=5.5mm]b.north east) node[midway, above=2pt] {30};
\draw[brc] ([yshift=10mm]a.north west) -- ([yshift=10mm]c.north east) node[midway, above=2pt] {50};
\draw[brc] ([yshift=14.5mm]a.north west) -- ([yshift=14.5mm]d.north east) node[midway, above=2pt] {100 available};
\node[below=1mm of a, align=center, text width=1.8cm] {5 folds:\\6 train / 2 val\\/ 2 test};
\node[below=1mm of c, align=center, text width=3.2cm] {added to training only};
\node[below=1mm of e, align=center, text width=2.1cm] {external test,\\inference only};
\end{tikzpicture}}
\caption{Use of the LH realizations in each suite. Nested training sets of 10, 30, 50 and 100 available realizations share the same five folds over LH0--LH9, whose validation and test boxes never change. LH100--LH199 are used only once, to evaluate the frozen 100-LH models. Segment widths are schematic.}
\label{fig:splits}
\end{figure}
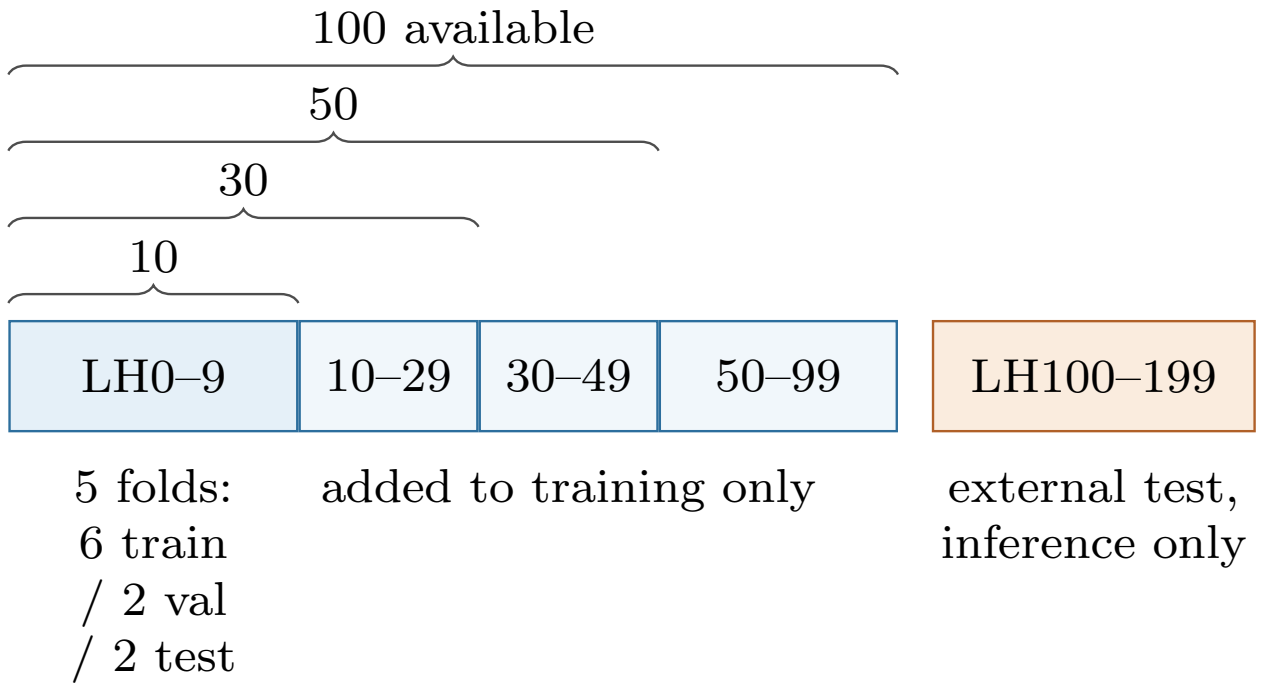

The splits are summarised in Fig.~\ref{fig:splits} and Table~\ref{tab:samples}. For the learning curves we keep the five folds previously defined over LH0--LH9. In each fold six of these ten realizations are used for training, two for validation and two for the internal test panel. At available-set sizes of 10, 30, 50 and 100, the additional realizations LH10--LH29, LH10--LH49 or LH10--LH99 enter training only. The number of source-training graphs per fold is therefore 6, 26, 46 or 96, while the validation and internal test graphs stay the same at every size. This design separates the effect of more diverse source-training simulations from any change in the evaluation sample.

The external test panel is distinct. It consists of LH100--LH199 for each of the three target suites, that is, 100 new simulations per suite. None of these realizations is used to train, validate or select the frozen 100-LH checkpoints. All five source-fold models are evaluated on the same external boxes, and their per-box metrics are averaged for the primary presentation; they are not treated as five independent test samples. The external panel contains 300 distinct boxes and, in the saved catalogue selection, 174,487 galaxies. In 18 of these boxes no selected galaxy lies in a void.

\begin{table}
\caption{Samples used in the two evaluation experiments, per suite.}
\label{tab:samples}
\centering
\small
\begin{tabular}{lcc}
\hline
 & Learning curve & External test \\
\hline
Training boxes per fold & 6, 26, 46, 96 & 96 (frozen) \\
Validation boxes per fold & 2 & -- \\
Test boxes & 2/fold (LH0--9) & 100 (LH100--199) \\
Source $\times$ target pairs & 9 & 9 \\
\hline
\end{tabular}
\end{table}

Each LH realization differs from the others in its cosmological and astrophysical parameters as well as in its initial conditions. Cross-suite differences therefore cannot be read as the effect of a single feedback prescription without further controlled comparisons, and the parameter analyses below are observational within each suite.

\section{Models and evaluation methods}
\label{sec:methods}

\subsection{Architectures}
\label{subsec:models}

All six models map the ten standardized descriptors of each galaxy to four logits, one per tidal-web class. The baseline is a node-feature-only multilayer perceptron (MLP), which applies the same fully connected network to each galaxy's Delaunay-derived descriptors without inter-node message passing. It measures how much of the classification signal is available without exchanging information between neighbours. The other five models are message-passing graph neural networks. Each layer updates the hidden state $\mathbf{h}_i$ of galaxy $i$ from its own state and an aggregate over its Delaunay neighbours,
\begin{equation}
\mathbf{h}_i^{(l+1)}=U^{(l)}\!\left(\mathbf{h}_i^{(l)},\;\bigoplus_{j\in\mathcal{N}(i)} M^{(l)}\!\left(\mathbf{h}_i^{(l)},\mathbf{h}_j^{(l)},\mathbf{e}_{ij}\right)\right),
\label{eq:mp}
\end{equation}
where $\bigoplus$ is a permutation-invariant aggregation (sum, mean or maximum), $M$ forms messages, $U$ updates the state and $\mathbf{e}_{ij}$ holds the edge attributes. The models differ in their choice of $M$, $\bigoplus$ and $U$ (see Ref.~\cite{Kololgi2026}).

The graph convolutional network (GCN) \citep{KipfWelling2017} aggregates linearly transformed neighbour states with fixed weights set by the node degrees, $\mathbf{h}_i'=\sigma\big(\sum_{j\in\mathcal{N}(i)\cup\{i\}}(\hat k_i\hat k_j)^{-1/2}\,\mathsf{W}\mathbf{h}_j\big)$, where $\hat k$ counts the self-loop. The graph attention network (GAT) \citep{Velickovic2018,Brody2022} replaces these fixed weights with learned attention coefficients,
\begin{align}
z_{ij}&=\mathbf a^{\mathsf T}\mathrm{LeakyReLU}\!\left(
\mathsf W_i\mathbf h_i+\mathsf W_j\mathbf h_j\right),\nonumber\\
\alpha_{ij}&=\underset{j\in\mathcal N(i)\cup\{i\}}{\mathrm{softmax}}(z_{ij}),\nonumber\\
\mathbf h'_i&=\sigma\!\left(
\sum_{j\in\mathcal N(i)\cup\{i\}}
\alpha_{ij}\mathsf W_j\mathbf h_j\right),
\label{eq:gat}
\end{align}

The GAT baseline uses GATv2 attention without edge
attributes, while GAT+ includes the edge attributes
in its attention scores. The equations show a single
attention head for clarity.

GAT+ follows \citet{Kololgi2026}: it uses several attention heads, includes the edge attributes in the attention score, and adds residual connections and layer normalization. EdgeConv \citep{Wang2019DGCNN} forms messages from a galaxy's state and its difference to each neighbour, 
$\mathbf{h}_i'=\max_{j\in\mathcal{N}(i)}\mathrm{MLP}\big([\mathbf{h}_i\,\Vert\,\mathbf{h}_j-\mathbf{h}_i\,\Vert\,\mathbf{e}_{ij}]\big)$, which makes it sensitive to local gradients of the features. The local graph Transformer \citep{Shi2021UniMP} uses dot-product attention restricted to Delaunay neighbours, with the edge attributes entering both the keys and the values,
\begin{align}
\alpha_{ij}&=\underset{j\in\mathcal{N}(i)}{\mathrm{softmax}}\!\left(\frac{(\mathsf{W}_q\mathbf{h}_i)^{\mathsf T}(\mathsf{W}_k\mathbf{h}_j+\mathsf{W}_e\mathbf{e}_{ij})}{\sqrt{d}}\right),\nonumber\\
\mathbf{m}_i&=\sum_{j\in\mathcal{N}(i)}\alpha_{ij}\big(\mathsf{W}_v\mathbf{h}_j+\mathsf{W}_e\mathbf{e}_{ij}\big),\nonumber\\
\mathbf{h}_i'&=\beta_i\mathsf{W}_r\mathbf{h}_i+(1-\beta_i)\mathbf{m}_i,
\label{eq:transformer}
\end{align}
where $d$ is the head dimension and $\beta_i$ is a learned sigmoid gate computed from $\mathbf{m}_i$, the transformed root state $\mathsf{W}_r\mathbf{h}_i$, and their difference. The graph models use the common Delaunay connectivity and, where their existing implementations provide for it, the periodic edge lengths as edge attributes.

\begin{table}[!h]
\caption{Architectures of the six classifiers. All take ten
node descriptors and predict four T-web classes. The Transformer
also uses 0.10 attention dropout.}
\label{tab:architectures}
\centering
\small
\setlength{\tabcolsep}{3pt}
\begin{tabular}{lll}
\hline
Model & Configuration & Dropout \\
\hline
MLP & $10\to10\to10\to10\to4$ & -- \\
GCN & $10\to15\to15\to4$ & -- \\
EdgeConv & $64$, 2 EdgeConv blocks & 0.10 \\
GAT & $64$, 2 GATv2 layers & 0.00 \\
GAT+ & $80$, 4 hidden + 1 output GATv2 & 0.50 \\
Transformer & $80$, 3 TransformerConv & 0.35 \\
\hline
\end{tabular}
\end{table}

\subsection{Training and preprocessing}
\label{subsec:training}

Model definitions, graph features and hyperparameters are held fixed as the number of source-training realizations grows. For each training fold, class weights are computed from the labels of its training graphs alone: $w_c=N_{\rm train}/(4N_c)$ for $N_c>0$, where $N_c$ is the number of training galaxies in class $c$ and $N_{\rm train}=\sum_c N_c$. If a class is absent, its weight is set to zero. Training minimizes the class-weighted cross-entropy,
\begin{equation}
\mathcal{L}=-\frac{\sum_i w_{y_i}\ln p_{i,y_i}}{\sum_i w_{y_i}},
\label{eq:loss}
\end{equation}
where $y_i$ is the T-web label of galaxy $i$ and $p_{i,c}$ is its predicted probability for class $c$.

All models use Adam \citep{Kingma2015Adam} with initial learning rate $10^{-3}$, weight decay $10^{-4}$ and gradient-norm clipping at 1.0. The five non-Transformer models use batches of four complete realization graphs, up to 2000 epochs and early-stopping patience of 250 epochs. A ReduceLROnPlateau scheduler monitors source-validation loss, reducing the learning rate by a factor of 0.70 after 100 epochs without improvement, down to a minimum of $10^{-6}$. The graph Transformer uses batches of two graphs, up to 600 epochs, early-stopping patience of 75 and scheduler patience of 30 epochs; its remaining optimizer and scheduler settings are unchanged. The principal seed is 42, and the checkpoint with the lowest source-validation loss is retained. Thus the architectures do not share identical optimization budgets.

Several descriptors are strongly skewed. For each source fold, a positive shift and a Box--Cox transform are fitted to these descriptors using the source-training realizations only,
\begin{equation}
y=\begin{cases}\big[(x+c)^{\lambda}-1\big]/\lambda, & \lambda\neq0,\\ \ln(x+c), & \lambda=0,\end{cases}
\label{eq:boxcox}
\end{equation}
followed by standardization to zero mean and unit variance; \citet{Kololgi2026} applied the same transform to their graph metrics. The fitted shifts, exponents, means and variances are saved with each checkpoint and reused unchanged for validation and for every target suite, so no information from a target suite enters the preprocessing. Before inference, the code checks that all shifted inputs lie in the positive domain of the transform.

\subsection{Learning curves and transfer evaluation}
\label{subsec:transfer}

All nine combinations of the three source and three target suites are evaluated separately. A source-trained model is applied to the target boxes without updating its weights. For each test box we compute the per-class precision $P_c$ and recall $R_c$, the per-class score $F1_c=2P_cR_c/(P_c+R_c)$, the accuracy, the macro-averaged score
\begin{equation}
\mathrm{macro\text{-}F1}=\frac{1}{4}\sum_{c=1}^{4}F1_c,
\label{eq:macrof1}
\end{equation}
and the weighted F1, in which each class enters with the fraction of the box's galaxies that carry its label. Macro-F1 is the primary metric because it gives the four classes equal weight, which matters when void galaxies are scarce: a classifier that never predicts a void can still reach a high accuracy. The scoring convention for a class that is absent from a box is the same throughout. We use \texttt{sklearn.metrics.f1\_score} with \texttt{labels=[0,1,2,3]} and \texttt{zero\_division=0}. A class absent from both the true labels and predictions receives an F1 score of zero rather than being omitted from the four-class macro average. This convention also applies to the 18 of 300 external boxes with no selected void galaxies.

The learning curves are evaluated on the fixed LH0--LH9 panel, and the external LH100--LH199 panel is evaluated only for the final 100-LH models. These are different experiments, and we never pool their results into a single performance estimate. On the external panel, each of the five source-fold models is applied to every target box, giving five fold-specific scores per box for each source--target--architecture combination. The primary summary averages these five scores for each box and then averages over boxes. It is a mean of single-model scores; we do not average predicted probabilities into an ensemble.

\subsection{Clustered uncertainty and paired comparisons}
\label{subsec:uncertainty}

Galaxies in one box share its large-scale field, initial seed and parameters, so the independent statistical unit is the simulation realization. Our 95 per cent intervals come from a box-level bootstrap: target boxes are resampled with replacement, the fold-averaged scores of each resampled box are kept together, and the interval is read from the percentiles of the resampled means 
($B=2400$ for the architecture means and paired contrasts in Fig.~\ref{fig:clustered}). When two architectures are compared, their difference is formed box by box before resampling. Within a target suite $t$ we also compare a model trained on another suite $s$ with the model trained on $t$ itself, on the same box $b$, and define the paired transfer gap
\begin{equation}
G_{s\to t}(b)=\mathrm{macro\text{-}F1}_{s\to t}(b)-\mathrm{macro\text{-}F1}_{t\to t}(b).
\label{eq:gap}
\end{equation}
Because both terms are measured on the same galaxies, the gap separates the effect of the training suite from differences in test-box composition.

The five source-training folds overlap, and we do not treat them as independent training experiments. The bootstrap intervals therefore describe the variation over external test boxes for the saved models and training settings. They do not include the variance that would come from retraining with fresh random seeds.

\subsection{Physical-property correlations}
\label{subsec:physical_analysis}

For each external box we record the number of selected galaxies, the fractions of galaxies assigned to the four classes and summary statistics of the Delaunay graph (mean degree and the mean and 90th-percentile edge lengths). We relate box-level performance to these quantities with Spearman rank correlations computed separately within each target suite. The suite-specific LH parameter tables provide $\Omega_{\rm m}$, $\sigma_8$ and the four feedback parameters. Because the feedback parameters have different physical meanings in each suite (Section~\ref{subsec:camels}), we analyse them within their own suite and never compare them directly across suites.

Multiple comparisons within a family of parameter tests are controlled with the Benjamini--Hochberg false-discovery-rate procedure \citep{BenjaminiHochberg1995}. For $m$ tests with ordered $p$-values $p_{(1)}\le\dots\le p_{(m)}$, the adjusted values are 
$q_{(i)}=\min\{1,\min_{j\ge i}m\,p_{(j)}/j\}$, and we report associations with $q<0.05$ within three separate families of six parameter tests, one per target suite (18 tests in total); the analysis of paired transfer gaps in Appendix~\ref{app:exploratory} uses one six-parameter family per source--target direction (36 tests in total). As an exploratory sensitivity check, the rank associations with CAMELS parameters are recomputed after adjusting for the logarithm of the galaxy count and the void fraction by ranking each CAMELS parameter, macro-F1 and both controls within each target suite, regressing the parameter and macro-F1 ranks separately on an intercept and the two ranked controls, and computing the Pearson correlation of the resulting residuals. This conditional analysis is not a causal model, since the simulation parameters influence both controls and the galaxy selection.

\section{Results}
\label{sec:analysis}

\subsection{Increasing the training sample}
\label{subsec:learning_results}

\begin{figure*}
\centering
\includegraphics[width=0.93\textwidth]{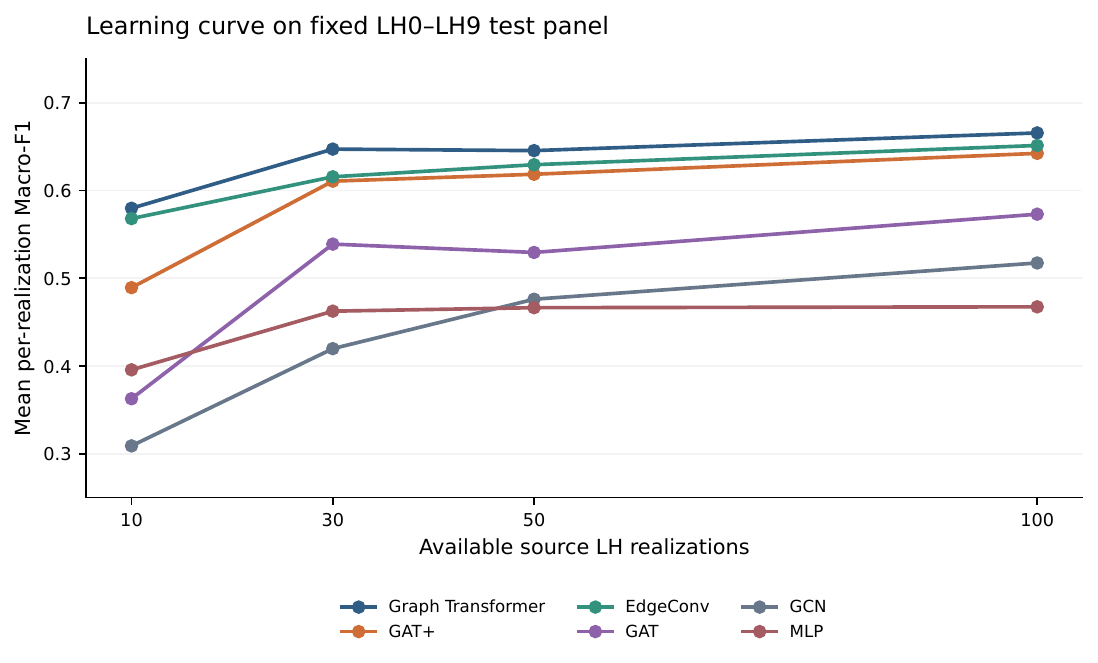}
\caption{Macro-averaged F1 as a function of the number of available source LH realizations, using the fixed LH0--LH9 validation and test assignments. Each point averages the three source and three target suites. At available-set sizes of 10, 30, 50 and 100, the source-training set contains 6, 26, 46 and 96 realizations per fold. The external LH100--LH199 panel is not used in this figure.}
\label{fig:learning}
\end{figure*}

Figure~\ref{fig:learning} shows the mean macro-F1 over the nine source--target combinations for all six models on the fixed internal test panel. With 100 available LH realizations, the Transformer, EdgeConv and GAT+ reach $0.666$, $0.651$ and $0.642$. With ten available realizations, their scores are $0.580$, $0.568$ and $0.489$. GAT and GCN gain the most, rising from $0.363$ to $0.573$ and from $0.309$ to $0.517$. The MLP improves from $0.396$ to $0.467$ but changes little beyond 30 available realizations. Not every curve is monotonic, yet all five graph models gain from additional source-training boxes, and for every model the largest step occurs between 10 and 30 available realizations (6 and 26 training boxes per fold). The MLP, which sees no neighbours, saturates early, so the extra simulations mainly help models that learn from the arrangement of neighbouring galaxies.

\subsection{Generalization to 300 unseen simulation boxes}
\label{subsec:external_results}

\begin{figure*}
\centering
\includegraphics[width=0.94\textwidth]{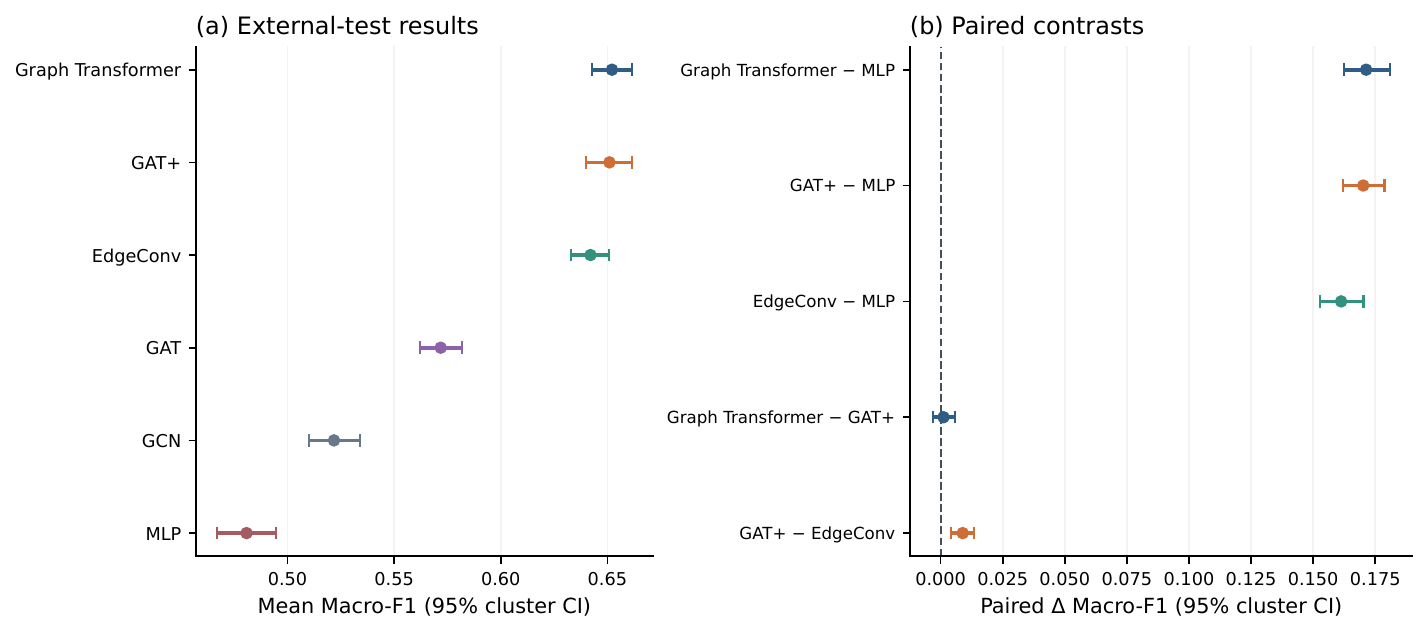}
\caption{(a) Mean external-test macro-F1 of the six models with 95 per cent box-clustered bootstrap intervals. (b) Paired differences in macro-F1, formed box by box before resampling. The dashed line marks zero difference.}
\label{fig:clustered}
\end{figure*}

The main generalization test is the separate LH100--LH199 panel. Table~\ref{tab:models} summarizes performance over the nine transfer directions. The Graph Transformer, GAT+ and EdgeConv reach mean macro-F1 values of $0.6522$, $0.6510$ and $0.6421$, against $0.4808$ for the MLP, with mean accuracies of $70.43$, $71.02$ and $69.24$ per cent. These external scores are close to the internal-panel values at 100 realizations (Section~\ref{subsec:learning_results}), so the leading models lose little when they move from the original ten test boxes to 100 new boxes per suite. Under the box-clustered bootstrap the Transformer macro-F1 interval is $0.6426$--$0.6613$. Its paired advantage over the MLP is $0.1714$ (95 per cent interval $0.1626$--$0.1808$). The difference between the Transformer and GAT+ is only $0.0012$, with an interval from $-0.0032$ to $0.0057$, so these two models cannot be ranked. Figure~\ref{fig:clustered} shows all six means with their clustered intervals and the main paired contrasts. All three attention or edge-convolution models are separated from the MLP by about $0.16$--$0.17$ in macro-F1, a gap far larger than any interval, whereas the leading three differ from each other by no more than $0.0101$; specifically, GAT+ exceeds EdgeConv by $0.00890$ (95 per cent paired box-bootstrap interval $0.00435$--$0.01334$).

\begin{table*}
\centering
\caption{Independent LH100--LH199 performance, averaged equally over the nine source--target directions and the 100 target realizations in each direction. Confidence intervals quoted in the text are box-clustered, not galaxy-level.}
\label{tab:models}
\begin{tabular}{lrrr}
\hline\hline
Model & Accuracy (per cent) & Macro-F1 & Weighted F1 \\
\hline
MLP & 52.56 & 0.4808 & 0.5332 \\
GCN & 58.48 & 0.5218 & 0.5866 \\
GAT & 62.95 & 0.5718 & 0.6311 \\
EdgeConv & 69.24 & 0.6421 & 0.6997 \\
GAT+ & 71.02 & 0.6510 & 0.7113 \\
Graph Transformer & 70.43 & 0.6522 & 0.7080 \\
\hline
\end{tabular}
\end{table*}

\begin{figure*}
\centering
\includegraphics[width=0.94\textwidth]{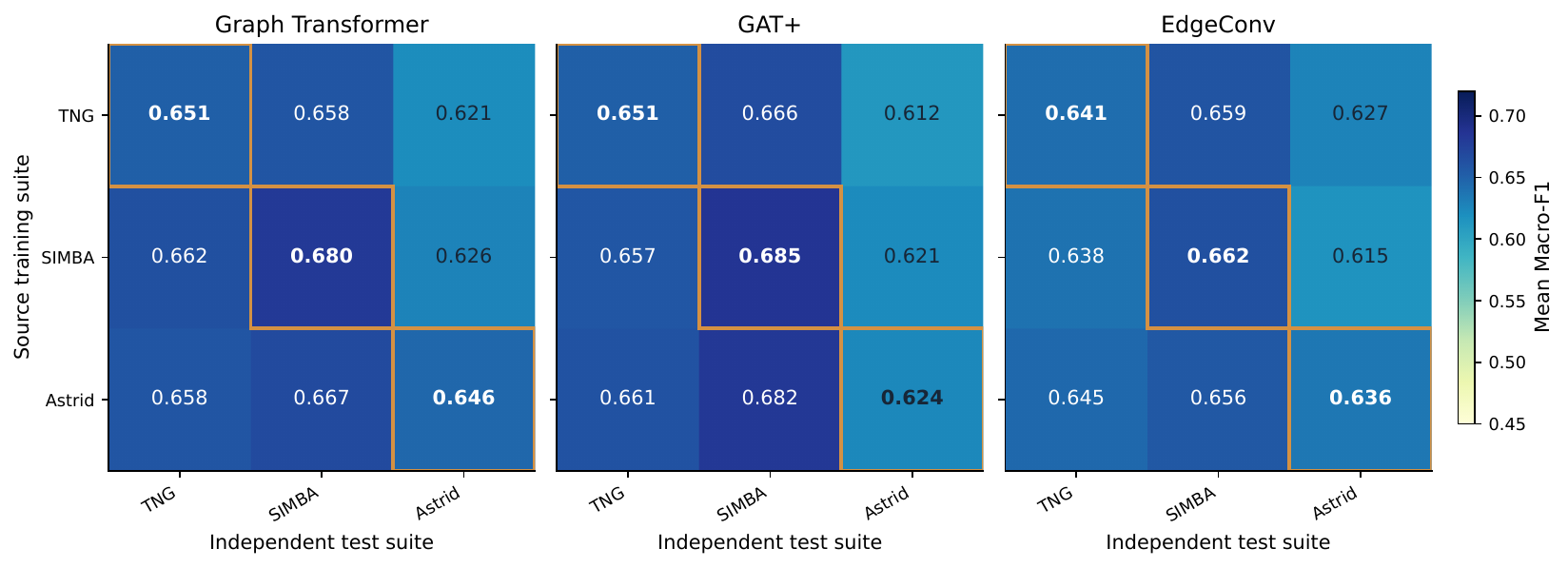}
\caption{Cross-suite macro-F1 matrices on the independent LH100--LH199 holdout for the three leading models. Rows give the source-training suite and columns the target-test suite. The five frozen source-fold models are applied individually to each external realization, and each entry is the mean of their box-level scores. Diagonal entries are same-suite results on independent realizations.}
\label{fig:transfer}
\end{figure*}

Figure~\ref{fig:transfer} shows that transfer depends on direction. For the Transformer, the same-suite diagonal ranges from $0.646$ on Astrid to $0.680$ on SIMBA, and the lowest off-diagonal entry is the TNG-trained model applied to Astrid ($0.621$). A model trained on another suite is not always worse: on the TNG holdout boxes, the SIMBA-trained Transformer exceeds the TNG-trained reference by $0.0114$ in macro-F1. On Astrid boxes, by contrast, the TNG-trained Transformer is lower than the Astrid-trained reference by $0.0252$. These differences should be read together with the suite-dependent galaxy populations (Section~\ref{subsec:richness_results}); they do not isolate the effect of one feedback prescription.

\subsection{Performance by tidal-web environment}
\label{subsec:class_results}

\begin{figure*}
\centering
\includegraphics[width=0.94\textwidth]{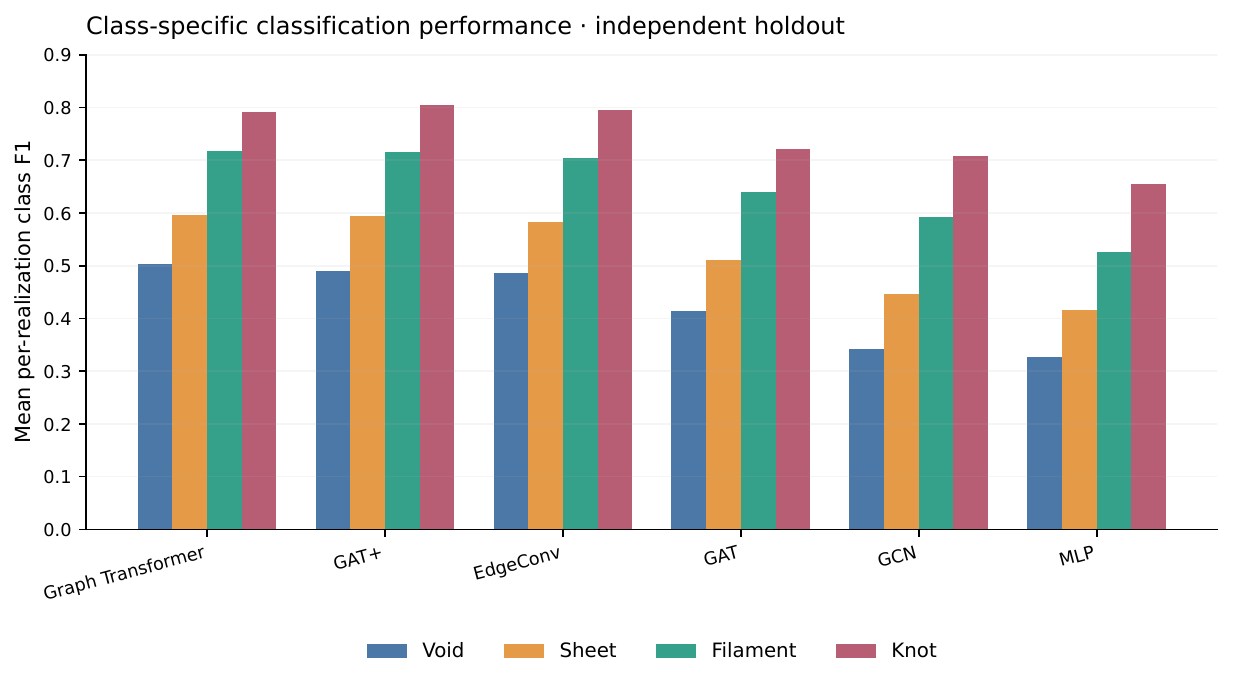}
\caption{External-test F1 by tidal-web environment for the six models, averaged over the LH100--LH199 panel. The numbers of selected galaxies differ between classes and suites; the handling of absent classes is described in Section~\ref{subsec:transfer}.}
\label{fig:classes}
\end{figure*}

Performance differs strongly between classes (Fig.~\ref{fig:classes}). On the external panel the Transformer reaches mean F1 values of about $0.503$, $0.596$, $0.717$ and $0.792$ for voids, sheets, filaments and knots. The ordering is the same for every architecture: knots are easiest and voids hardest. Void galaxies are rare above the stellar-mass threshold, and in 18 of the 300 external boxes no selected galaxy lies in a void. Such boxes make the treatment of absent classes in per-box macro averages important (Section~\ref{subsec:transfer}).

\subsection{Galaxy sampling and Delaunay geometry}
\label{subsec:richness_results}

\begin{figure*}
\centering
\includegraphics[width=0.94\textwidth]{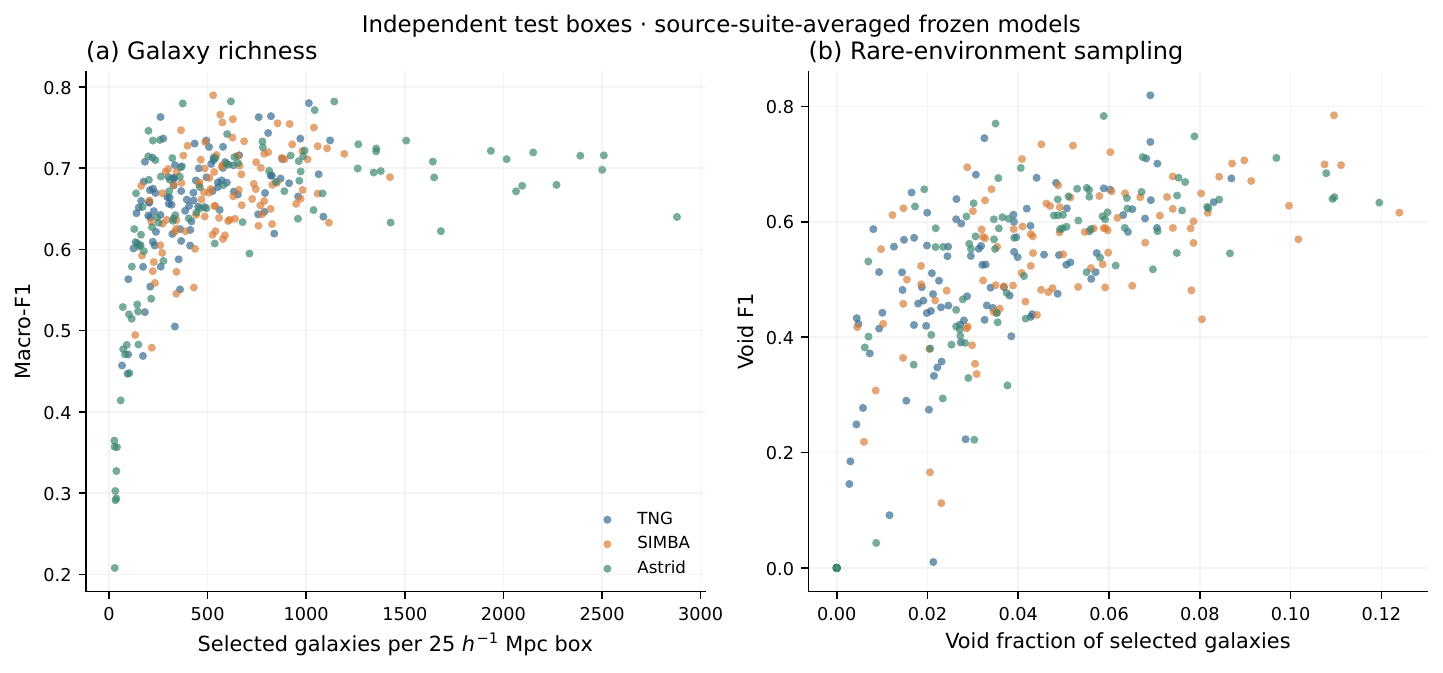}
\caption{(a) Box-level macro-F1 of the source-averaged frozen Transformer against the number of selected galaxies in each $25\,h^{-1}\mathrm{Mpc}$ external box. (b) Void F1 against the void fraction of the selected galaxies. Colours mark the target suite. The trends are descriptive and do not measure a causal effect of galaxy density.}
\label{fig:richness}
\end{figure*}

The number of selected galaxies per box correlates positively with the source-averaged Transformer macro-F1 in all three target suites (Fig.~\ref{fig:richness}a). The within-suite Spearman coefficients are $\rho=0.550$ for TNG, $0.448$ for SIMBA and $0.649$ for Astrid. The lowest scores, down to macro-F1 $\approx0.2$, occur in boxes with fewer than about a hundred selected galaxies, and above a few hundred galaxies the scores level off. Void F1 grows with the fraction of selected galaxies that lie in voids (Fig.~\ref{fig:richness}b): when only a handful of void galaxies exist, the class is hard to learn and noisy to score.

\begin{figure*}
\centering
\includegraphics[width=0.94\textwidth]{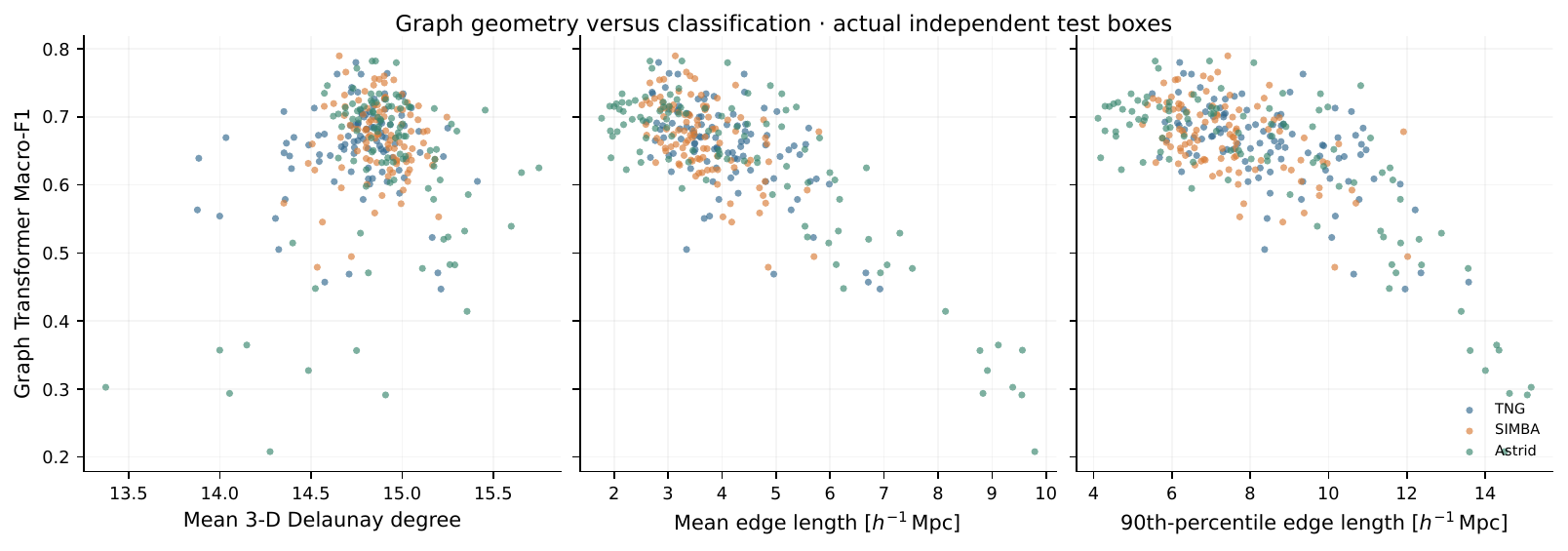}
\caption{Graph Transformer macro-F1 on the external boxes against three summaries of the Delaunay graph: mean three-dimensional degree, mean edge length and 90th-percentile edge length. Longer characteristic edges go with lower macro-F1; the trend partly reflects differences in the number of selected galaxies.}
\label{fig:geometry}
\end{figure*}

The mean periodic Delaunay edge length correlates negatively with macro-F1, with $\rho=-0.471$, $-0.515$ and $-0.661$ for TNG, SIMBA and Astrid (Fig.~\ref{fig:geometry}). The 90th-percentile edge length shows a similar trend, while the mean degree varies little between boxes. In a Delaunay graph the edge length scales roughly as the inverse cube root of the tracer density, so richness and edge length carry much of the same information. Both patterns point to tracer sampling as an important factor, but neither establishes a causal effect.

\subsection{Paired source-to-target transfer gaps}
\label{subsec:gap_results}

\begin{figure*}
\centering
\includegraphics[width=0.94\textwidth]{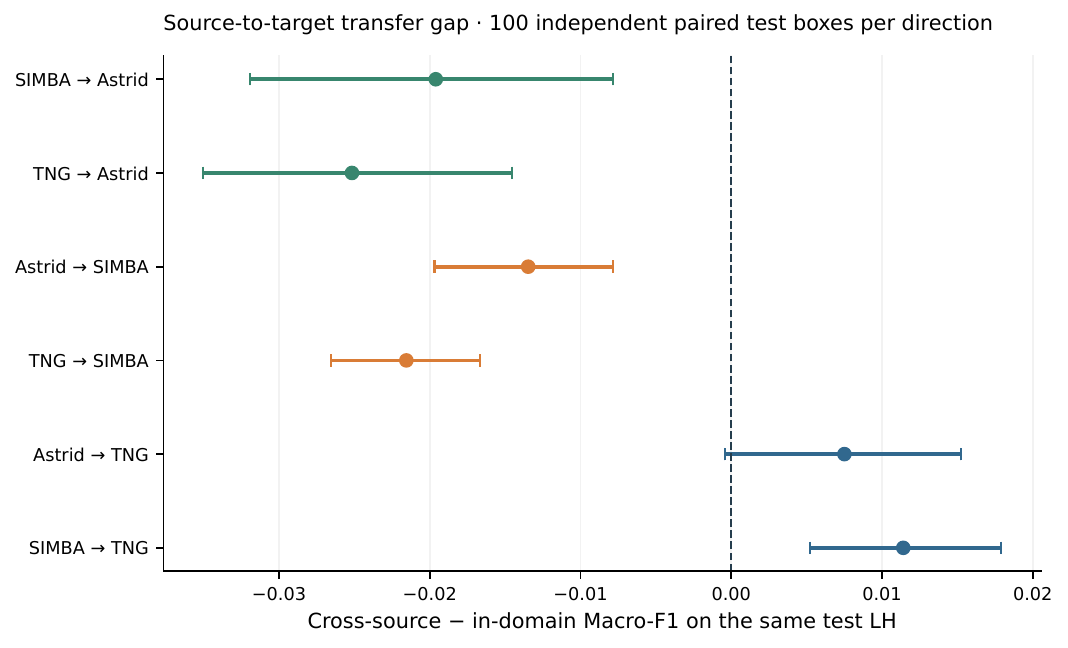}
\caption{Paired transfer gap $G_{s\to t}$ (Eq.~\ref{eq:gap}): macro-F1 of a Transformer trained on suite $s$ minus that of the Transformer trained on the target suite $t$, computed on exactly the same external boxes (100 per direction). Intervals resample independent test realizations; they are conditional on the frozen models and do not include retraining variance.}
\label{fig:gaps}
\end{figure*}

The paired gaps of Eq.~(\ref{eq:gap}) compare source suites on identical target boxes (Fig.~\ref{fig:gaps}). For the Transformer, the TNG-to-SIMBA gap is $-0.0216$ (95 per cent interval $-0.0265$ to $-0.0166$) and the SIMBA-to-TNG gap is $+0.0114$ ($0.0052$ to $0.0179$). The four directions that end on SIMBA or Astrid have negative gaps, whereas both directions that end on TNG are close to zero or positive. All gaps are small compared with the $\sim0.17$ 
advantage of the three leading graph models over the MLP. The choice of source suite matters, but no single suite is best for every target.

\subsection{CAMELS parameter associations}
\label{subsec:parameter_results}

\begin{figure*}
\centering
\includegraphics[width=0.94\textwidth]{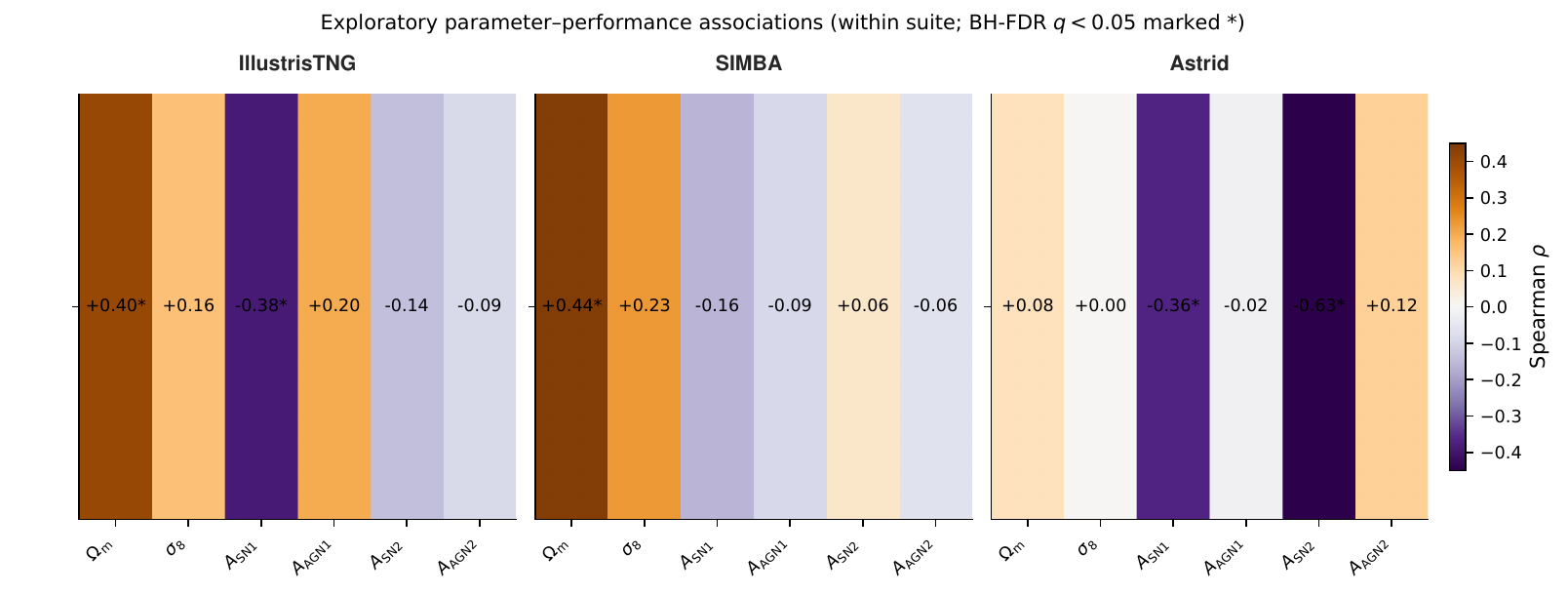}
\caption{Within-suite Spearman correlations between box-level macro-F1 and the CAMELS cosmological and feedback parameters, before adjustment. Asterisks mark associations with Benjamini--Hochberg $q<0.05$. Feedback parameter definitions are suite specific, so feedback-parameter correlations are not directly comparable across suites.}
\label{fig:parameters}
\end{figure*}

Some raw correlations between classification performance and the CAMELS input parameters are strong (Fig.~\ref{fig:parameters}). The unadjusted rank correlation with $A_{\rm SN2}$ in Astrid is $\rho=-0.627$, and the correlation with $\Omega_{\rm m}$ is $0.442$ in SIMBA and $0.403$ in TNG. Five of the 18 within-suite parameter tests pass the false-discovery-rate criterion before adjustment. After controlling for galaxy count and void fraction in the exploratory partial-rank analysis, none of the 18 adjusted tests has $q<0.05$. This attenuation is compatible with sampling and environment composition carrying much of the raw association. It does not demonstrate mediation, nor does it exclude a direct effect of the parameters, because the simulation parameters also shape both controls. The strongest raw association, and the parameter dependence of the paired transfer gaps, are examined in Appendix~\ref{app:exploratory}.

\section{Discussion}
\label{sec:discussion}

\subsection{What the graph adds, and what more simulations add}
\label{subsec:disc_graph}

The experiments support two conclusions. A graph representation carries information about the dark-matter web that a node-feature-only MLP does not recover, and a wider range of source-training realizations improves classification for most of the tested architectures. The first result concerns the value of neighbourhood structure. The MLP already receives each galaxy's own Delaunay descriptors, which include edge lengths and local shape, yet it trails the leading graph models by about $0.17$ in external macro-F1. Message passing lets a model combine the descriptors of a galaxy's neighbours, which is the information needed to tell, for example, a galaxy in a thin filament from one in a sheet with similar local density. The second result concerns the range of simulated galaxy configurations over which that representation stays useful. The MLP stops improving after about 30 available realizations, whereas GAT and GCN keep improving up to 100. Models that learn from the arrangement of neighbours need more examples of that arrangement.

The external holdout matters because the learning-curve test panel contains only the ten original realizations per suite. Those results show the response to more training data under a controlled evaluation set, but on their own they cannot show performance on fresh cosmological boxes. The LH100--LH199 panel supplies that evidence. On 300 unseen boxes the leading models keep mean macro-F1 values close to their internal-panel scores, although the nine transfer directions behave differently.

\subsection{Galaxy sampling as an organizing variable}
\label{subsec:disc_sampling}

Galaxy sampling appears to organize much of the variation. In boxes with more selected tracers, the Delaunay graph samples the web more densely and has shorter edges, and classification in those boxes is better. This fits the picture of a more informative geometrical description, but it is not proof that sampling density is the only limiting factor. Galaxy richness, class fractions, feedback parameters and the underlying matter field depend on one another, and the present tests cannot fully disentangle them. The raw CAMELS parameter correlations show the problem: several strong associations weaken once tracer count and void fraction are accounted for.

The $\Omega_{\rm m}$ dependence deserves a specific remark. Higher $\Omega_{\rm m}$ is expected to give more collapsed structure and more galaxies above a fixed stellar-mass threshold, which would raise performance through sampling alone. A second effect is specific to our labels. The T-web is computed from the cold-dark-matter density, and because $\Omega_{\rm b}=0.049$ is fixed while $\Omega_{\rm m}$ varies from $0.1$ to $0.5$ \citep{VillaescusaNavarro2021}, baryons make up about half of the matter at the low end of the range and about a tenth at the high end. The CDM field is therefore a less complete proxy for the total-matter tidal field at low $\Omega_{\rm m}$. Whether this contributes to the observed associations could be tested by recomputing the labels from the total matter density.

\subsection{Comparison with previous work}
\label{subsec:disc_previous}

\citet{Kololgi2026} reported an accuracy of 85 per cent and a macro-averaged F1 of $0.81$ for GAT+ with $M_\star>10^9\,M_\odot$ in IllustrisTNG-300. \citet{SuarezPerez2021} obtained 74 per cent accuracy and a weighted F1 of $0.728$ from $\beta$-skeleton features with a random forest. Our external accuracies of about 70 per cent are lower, and a direct comparison is limited for several reasons. Their samples come from a single large volume and are split at the galaxy level, so training and test galaxies share the same large-scale modes. Our test boxes are entire realizations with different seeds and parameters, and in six of the nine directions the test suite also differs from the training suite. The CAMELS boxes are also small: a $25\,h^{-1}\mathrm{Mpc}$ volume holds few large voids, and void galaxies, the class on which all methods do worst, are correspondingly scarce. The distinguishing contribution of this work is to quantify the reliability of galaxy-based tidal-web inference across hydrodynamical simulation suites, combining training-set scaling with independent cross-suite evaluation. We introduce neither the Delaunay graph nor the tidal-web classifier.

Our transfer results fit with earlier tests of GNNs across CAMELS suites. GNNs that infer halo masses or $\Omega_{\rm m}$ from galaxy catalogues lose some accuracy when the subgrid model changes \citep{VillanuevaDomingo2022,VillanuevaDomingo2022Graphs}, and domain adaptation can recover part of the loss \citep{Roncoli2023}. For the per-galaxy classification studied here the cross-suite penalties are small, between about $0.01$ and $0.03$ in macro-F1, and in two directions the cross-trained model matches or exceeds the in-domain one. One plausible reason is that the label depends on the matter field smoothed over $2\,h^{-1}\mathrm{Mpc}$, which gravity sets in much the same way in all three suites, so the galaxy-formation model acts mainly through which haloes host galaxies above the mass threshold. \citet{Ni2023} found that the wider spread of galaxy populations in Astrid helps models trained on it extrapolate to other suites. In our paired gaps the Astrid-trained Transformer loses little on TNG, but the SIMBA-trained model transfers to TNG at least as well, so our data do not single out one training suite.

\subsection{Limitations}
\label{subsec:disc_limits}

In this study, all analyses are performed at $z=0$ in relatively small simulation boxes with side lengths of $25\,h^{-1}\mathrm{Mpc}$. The threshold $M_\star>10^9\,M_\odot$ selects different galaxy populations across suites, and some boxes contain no selected void galaxies. The five source-fold training sets overlap, and the main results are based on a single training seed; therefore, the bootstrap intervals do not capture retraining variance. The Transformer and other models were also trained with different epoch budgets. The labels rely on a single smoothing scale and eigenvalue threshold, while the analysis uses real-space galaxy positions without survey selection effects or redshift-space distortions. Although the external test on 100 realizations per suite provides a substantially stronger assessment than the initial ten-realization evaluation, it does not replace validation against observational data or simulations beyond CAMELS.

\subsection{Future tests}
\label{subsec:disc_future}

The parameter analysis is a first characterization of the conditions that accompany good or poor transfer. It is not a causal measurement of galaxy-formation physics. Several tests follow directly. Selecting galaxies at a fixed number density, instead of a fixed stellar mass, would separate tracer sampling from the galaxy-formation model. Repeating the analysis at several stellar-mass thresholds and redshifts would show how the conclusions scale with sampling. Measuring errors as a function of the distance of each galaxy's eigenvalues from $\lambda_{\rm th}$ would show how many errors fall on physically ambiguous boundaries; \citet{Kololgi2026} suggested replacing the discrete labels with a regression on the continuous eigenvalues for the same reason. Labels from the velocity-shear web \citep{Hoffman2012} or from a topological method applied to the matter field \citep{Sousbie2011} would test whether the conclusions depend on the web definition, and domain-adaptation losses \citep{Roncoli2023} could reduce the remaining cross-suite gaps. These experiments are outside the present frozen benchmark, and we make no claim based on any of them.

\section{Conclusions}
\label{sec:conclusion}

We have investigated how reliably the spatial geometry of galaxies predicts their dark-matter tidal-web environments across three CAMELS hydrodynamical suites, using six neural-network architectures and a common set of periodic Delaunay descriptors. Our main findings are as follows.

\begin{enumerate}
\item Increasing the number of available source-training realizations from 10 to 100 improves the performance of graph-based models, with the largest gains occurring between 10 and 30 realizations. The MLP, which uses Delaunay-derived descriptors without message passing, shows comparatively little improvement beyond 30 realizations.
\item An inference-only evaluation on LH100–LH199 shows that the leading models generalize well to 300 previously unseen simulation volumes. Averaged over all nine source-to-target directions, the Graph Transformer and GAT+ achieve external macro-F1 scores of approximately $0.65$, compared with $0.48$ for the MLP. The difference between the two leading models is not statistically significant.
\item Classification errors are not uniform across environments. Voids are more difficult to identify than sheets, filaments, or knots. Paired transfer gaps are generally small in magnitude, around $0.01–0.03$ in macro-F1, and depend on the source and target suites. For TNG targets, models trained on SIMBA or Astrid perform comparably to the TNG-trained model.
\item Box-level performance is associated with galaxy abundance and Delaunay edge length. Several raw associations with CAMELS cosmological and feedback parameters weaken after accounting for tracer count and void fraction.
\end{enumerate}

These results support the transferability of galaxy-graph-based tidal-web inference across the three CAMELS hydrodynamical suites, with classification performance closely associated with the abundance and spatial distribution of the selected galaxy tracers. Any physical interpretation of the remaining differences needs care, particularly when comparing galaxy-formation prescriptions that also change the abundance of the selected tracers.

\section*{Acknowledgements}
We thank the CAMELS collaboration for making their simulations, catalogues and gridded fields publicly available.

\section*{Data availability}
The CAMELS simulations, halo and subhalo catalogues and density grids used in this work are publicly available from the CAMELS project. The analysis uses the frozen \texttt{Web3} pipeline, the source-fold manifests, the fitted preprocessing records, the experiment outputs and the independent LH100--LH199 evaluation tables. These analysis products and the associated code will be made publicly available in a versioned repository upon publication. The raw simulation data are not redistributed with this paper.

\appendix
\section{Exploratory parameter analyses}
\label{app:exploratory}

The analyses in this appendix are exploratory. They were not specified before the external evaluation and should be read as descriptions of the saved results.

\begin{figure}
\centering
\includegraphics[width=\columnwidth]{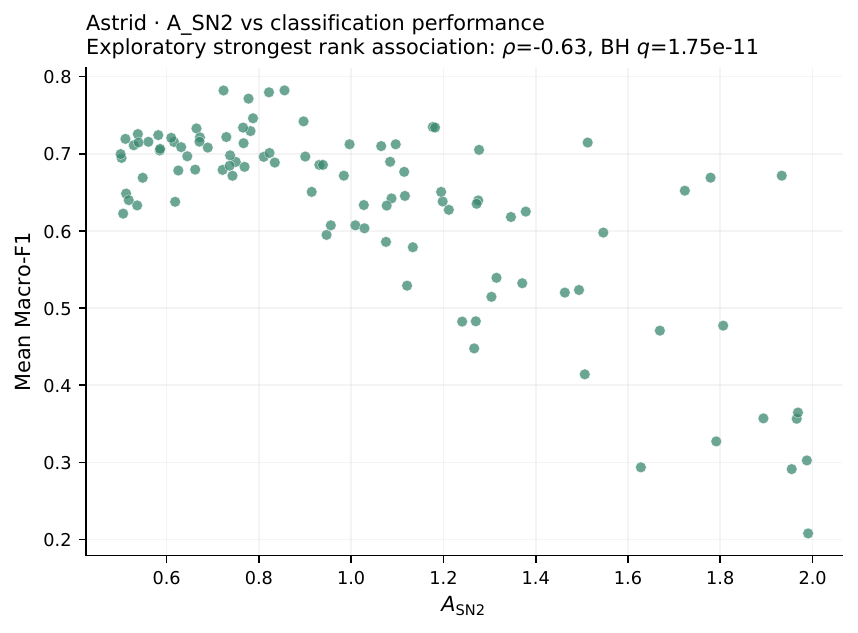}
\caption{Mean macro-F1 of the external Astrid boxes against $A_{\rm SN2}$, the strongest unadjusted association in Fig.~\ref{fig:parameters} ($\rho=-0.63$, Benjamini--Hochberg $q=1.75\times10^{-11}$).}
\label{fig:asn2}
\end{figure}

The strongest raw association between performance and a CAMELS parameter is with the Astrid supernova parameter $A_{\rm SN2}$, which scales the speed of the supernova-driven wind in that model \citep{Ni2023}. Figure~\ref{fig:asn2} shows the box-level scores against $A_{\rm SN2}$. For $A_{\rm SN2}\lesssim1.2$ most boxes reach macro-F1 between about $0.6$ and $0.8$, while the poorest boxes, with scores between about $0.2$ and $0.5$, cluster at the high end of the range. The supernova feedback in Astrid is energy driven, so a larger $A_{\rm SN2}$ gives faster winds with a lower mass-loading factor, $\eta_{\rm w}\propto v_{\rm w}^{-2}$ \citep{Ni2023}. 
Across the 100 independent Astrid test boxes, $A_{\rm SN2}$ is strongly anticorrelated with the number of selected galaxies (Spearman $\rho=-0.868$, $p\simeq1.6\times10^{-31}$). This is consistent with tracer abundance contributing to the unadjusted association between $A_{\rm SN2}$ and macro-F1, which is no longer significant after exploratory adjustment for galaxy count and void fraction. It does not establish a causal feedback mechanism.

\begin{figure}[!htbp]
    \centering
    \includegraphics[width=1.0\columnwidth]{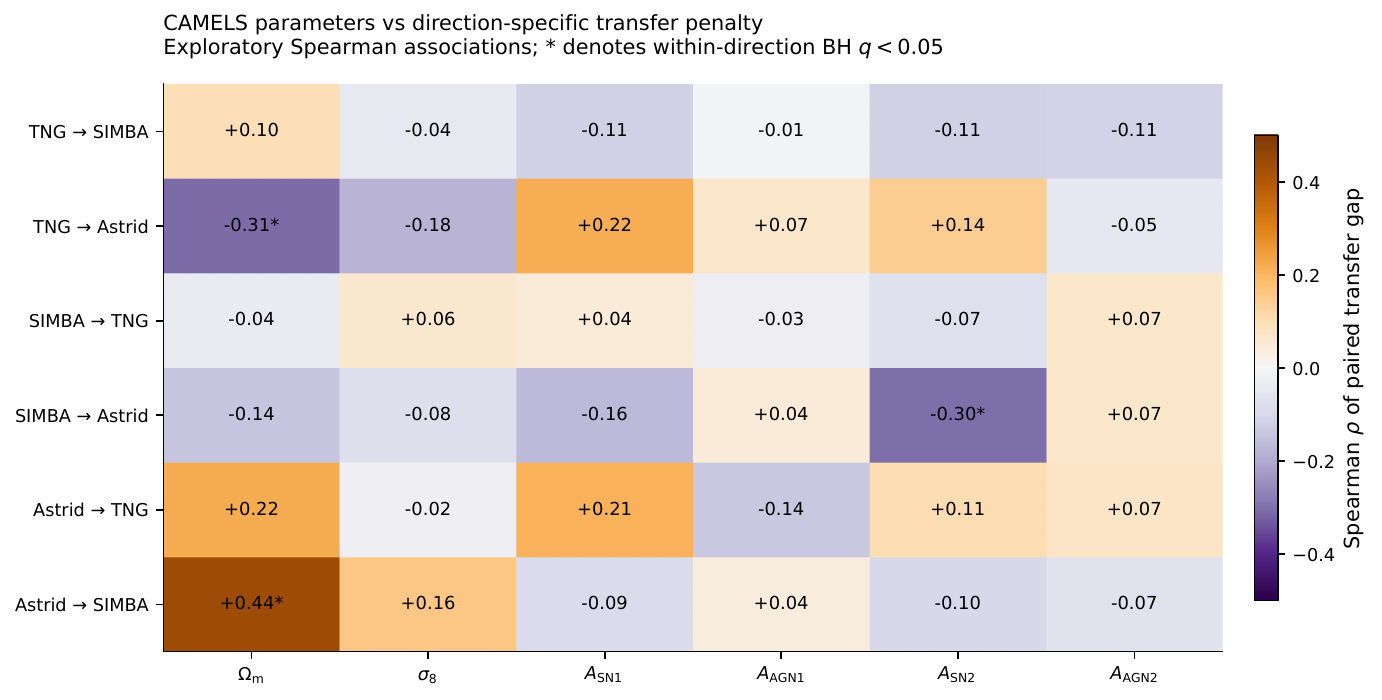}
    \caption{Spearman correlations between the CAMELS parameters
    of each external target box and the paired transfer gap for
    each source--target direction. Asterisks mark $q<0.05$
    after Benjamini--Hochberg correction within each direction.}
    \label{fig:gapparams}
\end{figure}

Figure~\ref{fig:gapparams} relates the paired transfer gaps of Eq.~(\ref{eq:gap}) to the CAMELS parameters of the target boxes. Three of the 36 direction--parameter tests have $q<0.05$ within their direction: the gap correlates with $\Omega_{\rm m}$ for Astrid-trained models on SIMBA ($\rho=+0.44$) and for TNG-trained models on Astrid ($\rho=-0.31$), and with $A_{\rm SN2}$ for SIMBA-trained models on Astrid ($\rho=-0.30$). Because the parameters of a target box also set its galaxy count and environment fractions, these correlations do not identify which physical difference between the suites causes the transfer penalty.

\bibliography{misc/reference}

\end{document}